\documentclass[aps,prl,preprint,amsmath,amsfonts,amssymb, braket, groupedaddress,floatfix]{revtex4-2}

\usepackage{graphicx}
\usepackage{dcolumn}
\usepackage{bm}
\usepackage{hyperref}
\usepackage[usenames,dvipsnames]{color}
\usepackage[normalem]{ulem}
\usepackage{amsmath}
\usepackage{amssymb}
\usepackage{braket}

\begin{document}


\title{Spontaneous Emission, Work Potential and Relaxation-Limited Processes in Setting Limits on Solar Energy Conversion Efficiency}

\author{Sumanta Mukherjee$^{1,*}$}

\affiliation{$^1$ Solid State and Structural Chemistry Unit, Indian Institute of Science, Bengaluru, Karnataka 560012, India\\}

\email{sm31081985@gmail.com}

\date{\today}

\begin{abstract}
Understanding the thermodynamics of radiation and the quantum-mechanical interactions between light and matter is important both for theoretical purposes and for technological advances, such as determining the limits of key processes like light-to-usable-energy conversion efficiencies. In this report, we discuss the physics of these two aspects, considering spontaneous emission as a pathway, and highlight the limitations of such descriptions in assessing energy-harvesting efficiency.  In view of these limitations, we adopt a simplified approach to evaluate the exergy and work potential of radiation, providing a framework for assessing various aspects of light-to-usable-energy conversion efficiency. Our approach allows a theoretical estimate of the thermodynamic maximum limit for light-to-usable-energy conversion, which is approximately 76$\%$. We validate these exergy and work potential estimates by modeling and accurately reproducing the Shockley–Queisser limit ($\sim 33.3\%$), which imposes a practical constraint on solar-to-usable-energy conversion efficiency. Beyond exergy considerations, our model incorporates processes such as spontaneous emission, nonradiative thermal losses, and photon upconversion, allowing us to evaluate their roles. The model further suggests that, under certain conditions, the maximum conversion efficiency can reach approximately 48$\%$, for example with multijunction solar cells or via photon upconversion. These findings further suggest that the true thermodynamic limit for light-to-usable-energy conversion may be much higher (approximately 76$\%$). However, accurately estimating this limit requires a more complete understanding of the thermodynamics of light, light–matter interactions, and the connection between them.
\end{abstract}

\maketitle

\section{Introduction}

Spontaneous emission of light from an excited source\cite{1,2,3,4,5,6} has enabled a variety of major scientific breakthroughs, ranging from the physics of blackbody radiation\cite{6,7,8} and the emergence of quantum mechanics\cite{1,6} to the development of quantum field theories\cite{9,10,11}. While it is now generally understood that spontaneous emission originates from the interaction between the underlying electromagnetic field and the excited state of matter\cite{2,3,5,9,10,12}, several fundamental aspects of its theoretical description require further development. For example, what causes the spontaneity of spontaneous emission?\cite{13} Does radiation have a temperature?\cite{14,15,16,17,18} What is the entropy of radiation?\cite{17,19,20,21,22,23} While the second law of thermodynamics states that a spontaneous process must increase (or remain constant in the case of a completely reversible process) the entropy of the universe (system and surroundings)\cite{24,25}, is it possible to provide such a description for spontaneous emission?\cite{13} While these thermodynamic descriptions of electromagnetic fields\cite{15,20,26} and spontaneous emission of light are important for theoretical purposes, they are also significant for enabling technological advances through an improved understanding of phenomena such as light-to-usable-energy conversion efficiency in solar cell devices\cite{17,18,27}, energy conversion efficiency in photosynthesis\cite{16,28,29,30,31}, the efficiencies of lasers\cite{32,33,34,35} and light-emitting devices\cite{17,36,37}. Before describing the thermodynamic aspects of light, it is worth noting that the thermodynamics of light has been studied for a long time, with numerous attempts to characterize it by associating temperature\cite{15,17,18,38,39} and entropy flux\cite{17,40,41} with the radiation (e.g., blackbody radiation). In the following section, we generalize these approaches and pave the way for a better understanding of related phenomena. The present discussion demonstrates that the general thermodynamic description of the radiation \cite{17,40}, although widely employed \cite{26,29}, may be insufficient for understanding the origin of spontaneity and the associated phenomena of spontaneous emission. Rather, the observed spontaneity arises from a deeper quantum-mechanical understanding of light–matter interactions, as described by various theoretical approaches\cite{3,5,10,11,12}. We have presented a brief discussion of the well-known theories\cite{3,5,10,11,12} of light–matter interaction in the forthcoming section.\\
However, we find that even quantum-mechanical descriptions of light–matter interactions are not complete enough to capture various thermodynamic aspects of light. In view of such limitations, we provide a simplified estimate that allows us to evaluate certain thermodynamic properties of radiation, such as work potential and exergy\cite{24,25}. In particular, by accounting for the microscopic details of the energy transfer path, we provide a more realistic estimate of the work potential than the exergy estimate proposed by R. Petela, which is based on an idealized scenario \cite{42,43}. We justify our estimate by using an effective model to evaluate the solar-to-usable-energy conversion efficiency, known as the Shockley–Queisser (S–Q) limit \cite{44}. Alongside estimates of work potential, the presented model allows for the inclusion of various processes, such as spontaneous emission and nonradiative thermal losses, thereby elucidating their roles. The presented model allows us to express the approximate efficiency ($\eta_E$) of a single-junction solar cell as $\eta_E = 0.76 \times \mathcal{Q} \times \Lambda_{\mathrm{L}}$, where $\mathcal{Q}$ is defined as the ``\emph{quality factor}'' and $\Lambda_{\mathrm{L}}$ is defined as the ``\emph{loss-modified retention factor}''. Further, our simplified model effectively reproduces the $\sim$ 33.3$\%$ maximum efficiency under standard conditions, as predicted by the S–Q limit, thereby supporting the validity of the exergy estimates and the applicability of the presented model. Furthermore, the simplicity of the presented model allows us to effectively model different conditions, such as multi-junction solar cells or photon upconversion effects, and to estimate the conversion efficiencies under these conditions. Based on this description, we suggest that the maximum solar efficiency can exceed the S–Q limit \cite{44,45} ($\sim$ 33.3$\%$) and may reach higher efficiencies of approximately 48$\%$ under these circumstances. Such high values of efficiency have indeed been observed, for example, in multi-junction solar cells \cite{46,47} or through the reduction of thermal heat loss following photon absorption \cite{45}. Efficiencies higher than the S–Q limit have also been suggested by various theoretical calculations under different conditions, such as hot-carrier injection \cite{48} or multijunction solar cells \cite{49}. Considering all these effects, we therefore suggest that the maximum efficiencies of $\sim$ 33.3$\%$ to 48$\%$ estimated under different circumstances represent conditional efficiencies only, and that the true thermodynamic efficiency of solar-to-usable energy conversion is close to 76$\%$, as estimated by the exergy analysis presented here. However, accurately estimating this limit requires a more complete understanding of the thermodynamics of light, light-matter interactions, and the connection between them.

\section{Methodological Context}

\subsection {Thermodynamics of light}
First, let us start with the general quantum-mechanical description of a radiation field. Within the framework of quantum electrodynamics and the quantum description of the electromagnetic field, the Hamiltonian ($\hat{\mathcal{H}}$) of an electromagnetic field enclosed in a cubic box of length $L$ can be expressed as an infinite set of harmonic oscillators\cite{9,50}.
\begin{equation}
\hat{\mathcal{H}} = \sum\limits_{k} \hslash \omega_k \left( a^\dag _k a_k + \frac {1}{2} \right)
\end{equation}
Here, the sum is over all possible modes defined by $k$. $\hslash$ is the reduced Planck constant, and $\omega_k$ is the angular frequency. Note that, for simplicity, we have omitted the polarization (helicity) ($\gamma =  \pm 1$) of the modes in the above expression. $a^\dag _k$ and $a_k$  are the creation and annihilation operators, respectively.\\
Consider a beam of radiation (such as blackbody radiation or radiation from an LED source) occupying a certain number of modes with angular frequency $\omega_i$ and occupancy $n_i$ for the $i^{\text{ th}}$  mode. In such cases, considering the bosonic nature of the photon, the entropy change ($dS$), due to the emission of these photons, can be calculated as\cite{17,32,34}.
\begin{equation}
dS = k_B \sum\limits_{i} \left[ \left(1+ n_i\right) \ln \left(1+ n_i\right) - n_i \ln n_i \right]
\end{equation}
Here, $k_B$ is the Boltzmann constant. To account for additional degrees of freedom, for example due to the large density of states (DOS) around a particular frequency mode, the following expression for the entropy change may be used.\cite{17}.
\begin{equation}
dS = k_B \sum\limits_{i} F_i \left[ \left(1+ n_i\right) \ln \left(1+ n_i\right) - n_i \ln n_i \right]
\end{equation}
Here, $F_i$ is related to the degrees of freedom, or DOS for the present case, of the photons around the $i^{\text{ th}}$ mode with angular frequency $\omega_i$.\\
Considering that there is no volume change of the photons during emission, the first law of thermodynamics\cite{17} can be used to calculate the effective temperature of this photon distribution as $T_{eff} = dE/dS$ at constant volume\cite{17,34}, providing a possible estimate of the temperature of the emitted photons. Here $dE$ is the change in internal energy.
While it is tempting to relate this estimate of entropy and temperature to the emitted light, it is important to note that the photons under general conditions (for example, when generated by an LED source) may not exhibit an equilibrium distribution of occupation numbers given by the Bose–Einstein factor \cite{6,32} over all possible modes. Therefore, the situation described above does not represent an equilibrium thermodynamic temperature ($T_{th}$); instead, the estimated temperature can only be considered an effective temperature ($T_{eff}$). However, for an equilibrium radiation source-for example, a blackbody radiating photons distributed over all modes-the effective temperature may resemble a thermodynamic temperature of the radiation, given by the temperature of the blackbody itself ($T_{eff} = T_{th} = T_{BB}$)\cite{34}.\\
Considering all the factors mentioned above, the following two equations have been derived and widely used in various contexts for the energy and entropy fluxes ($\dot{E}$ and $\dot{S}$, respectively) of photons emitted from a blackbody at a given temperature. These equations are believed to have been first derived by Planck\cite{22,23} and generally take the following forms\cite{32,34,41}.
\begin{equation}
\begin{aligned}
&\dot{S} = \frac {2k_B}{c^2} \int_{A} \int_{\Omega} \int_{\Delta \nu}  B^\prime \nu^2 d\nu \cos \theta d\Omega dA \\
&\text{where } B^\prime = \left[ \left(1+ n\right) \ln \left(1+ n\right) - n \ln n \right]
\end{aligned}
\end{equation}
Here, $c$ is the speed of light, $n$ is the occupation number of a mode, and $\nu$ is the frequency. $A$, $\theta$ and $\Omega$ are related to the surface area and solid angle of the emissions. The expression for the energy flux is given by\cite{32,34}.
\begin{equation}
\dot{E} = \frac {2h}{c^2} \int_{A} \int_{\Omega} \int_{\Delta \nu}  n \nu^3 d\nu \cos \theta d\Omega dA
\end{equation}
Here, $h$ is the Planck constant. These equations, with some further calculations for blackbody emission, lead to the general expressions for the total energy flux density and the entropy flux density\cite{32,34}, which are given by:
\begin{equation}
\begin{aligned}
&E_{BB} \propto T_{BB}^4\\
&S_{BB} \propto T_{BB}^3
\end{aligned}
\end{equation}
This yields the temperature of the radiation to be equal to the temperature of the black body, $T_{BB}$. Since this radiation condition provides an equilibrium population distribution for all modes of radiation, the radiation temperature may be regarded as an equilibrium thermodynamic temperature($T_{eff} = T_{th} = T_{BB}$)\cite{34}. Considering this equilibrium condition and temperature, various thermodynamic parameters-for example, the exergy of radiation\cite{42,43,51} and Carnot efficiency\cite{17,18,28,29,39}-have been used to estimate the efficiency of different processes, such as photosynthesis\cite{16,28,29,30,31}. While the above discussion provides a brief overview of the entropy and temperature of photons, as presented in many studies, it is important to recognize that the spontaneity of emission and absorption processes has a deeper origin than that suggested by an effective radiation temperature, as will be discussed in the following section. In that section, we focus primarily on the entropy changes associated with photon absorption and emission \cite{52,53,54,55,56,57,58,59}, arising from the quantum-mechanical interaction between the photon field and the matter field \cite{3,5,10,11,12,60}.

\subsection {Light-matter interaction}
\begin{figure}[t]
\begin{center}
\includegraphics[width=1.0\columnwidth]{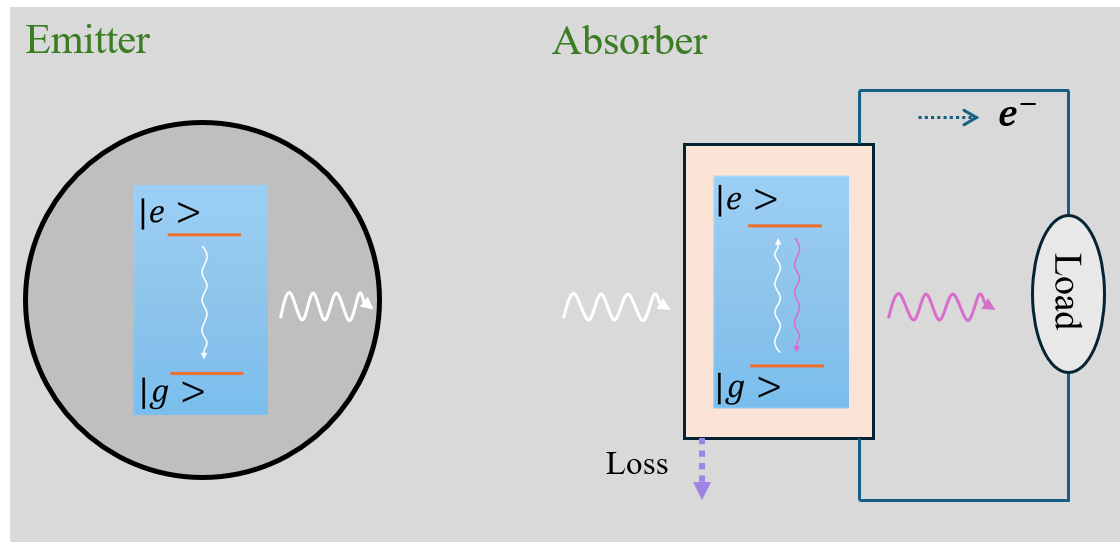}
\caption{This figure provides a conceptual schematic representation of spontaneous emission traveling from the emitter to the absorber and the subsequent methodologies (including exergy, and losses) used to derive the efficiency equation. A more detailed description is provided in the manuscript.}
\label{fig1}
\end{center}
\end{figure}
In the context of light–matter interaction, a large number of Hamiltonians\cite{61} have been used to model the underlying processes. For a closed system, the most representative Hamiltonian is the Jaynes–Cummings model\cite{53,55,58,60,62}, which explains phenomena such as Rabi oscillations\cite{53,55,60,62,63}, revivals\cite{53,55,58,64}, damping\cite{62,65,66,67,68}, and related effects in a lossless cavity. On the other hand, for open-system problems, which are more relevant to the present discussion, a system characterized by a ground state $\ket{g}$ and an excited state $\ket{e}$ interacts with a large number of photon modes. Such interactions lead to the damping of Rabi oscillations and their eventual suppression, resulting in the apparent irreversibility of the process\cite{5,10,63,65,67}. A model Hamiltonian\cite{5,62,63,69} describing this situation is given below.
\begin{equation}
\hat{\mathcal{H}_{t}} = \frac {\hslash \omega_0 \sigma_z}{2} + \sum\limits_{k} \hslash \omega_k a^\dag _k a_k + \sum\limits_{k} \hslash \text{ } \xi_k \left(  \sigma_{-} a^\dag _k + \text{ }  \sigma_{+} a_k  \right)
\end{equation}
Here, the ladder operators $\sigma_{+}$ and $\sigma_{-}$ take the atomic state to higher or lower energies, $\sigma_z$ is the Pauli matrix for a two-state system, and $\omega$ represents the angular frequency. The second term represents the electromagnetic field, shifting the zero-point energy, while the last term describes the interaction, where $\xi_k$ represents the coupling strength. The sum runs over all allowed photon emission modes, maintaining the conservation laws. The most commonly used approaches for explaining this irreversible behavior are the Weisskopf–Wigner theory\cite{10,70}, along with various Markovian\cite{11,71} and master-equation formalisms\cite{11,62,66,71,72}. However, without delving into the complexities of these theories, we highlight the key aspects that are relevant to and subsequently employed in the present model. Under the Born–Markov approximation\cite{11,62,71}, the interaction with the field modes is assumed to be weak, and the timescale over which bath correlations develop is presumed to be much shorter than the atomic decay time. Consequently, the bath is considered to possess no memory, resulting in Markovian dynamics. This approximation, together with the continuum of field modes\cite{10}, leads to the decay of the atomic population and provides an explanation for the apparent irreversibility of spontaneous emission\cite{10}. However, owing to the unitary evolution of the atom–field system, the total entropy remains conserved, although the entropy of its subsystems may vary with time\cite{73}. In contrast, within master-equation approaches, system–bath correlations are often neglected, and the non-unitary time evolution of the system's reduced density matrix is monitored\cite{71,72}. Consequently, the system entropy generally increases with time \cite{56,74,75}. In both the Weisskopf–Wigner theory and master-equation approaches to free-space emission, the initial field state is assumed to consist of vacuum modes at zero temperature, except in certain generalized cases\cite{10,71}. The spontaneity of these processes is commonly characterized using Spohn's method\cite{76}. While the above discussion provides a quantum-mechanical understanding of light–matter interaction and spontaneous emission, its connection to the classical thermodynamics discussed in the previous section of this manuscript appears to be largely unexplored, except for a few sporadic attempts\cite{77}. Therefore, we present a semiclassical description to investigate whether these two aspects of light are fundamentally connected. To this end, we consider the free-space regime, where multiple emitters interact with different vacuum modes and the interatomic separations are sufficiently large to prevent significant correlations among the emitters\cite{78}. Under these conditions, each emitter can be treated independently, as described by the Weisskopf–Wigner theory outlined above. Our aim is to provide physically meaningful estimates of the work potential and exergy available for light-to-electricity conversion in solar materials under normal operating conditions.

\section{Results and Discussion}

\subsection {Exergy of radiation}

Before estimating the free energy of radiation, we briefly review the derivation proposed by R. Petela \cite{42,43}, who estimated the exergy ($E_{ex}$) of sunlight within a classical thermodynamic framework. The resulting expression for the normalized exergy can be written approximately as follows:
\begin{equation}
E_{ex}/E = 1 - \frac {4}{3} \frac {T_0}{T_1} + \frac {1}{3} \left( \frac {T_0}{T_1} \right)^{4}
\end{equation}
where $T_1$ denotes the radiation temperature and $T_0$ denotes the environmental temperature. Such an estimate indicates that the exergy content of sunlight exceeds 90$\%$, implying that over 90$\%$ of its energy is theoretically available for conversion into useful work. However, we find that this derivation somewhat oversimplifies the quantum nature of radiation, which motivates us to investigate whether more accurate estimates of the work available from sunlight can be obtained. Before doing so, we rewrite the expressions for thermal blackbody radiation, such as sunlight, that serve as conversion factors in our estimation of the available work. For thermal blackbody radiation, the average energy can be estimated using the following expression \cite{6,44}.
\begin{equation}
\begin{aligned}
&\langle \epsilon_{av} \rangle = \frac {C_n  \int_{0}^{\infty} h\nu^3 \frac {1} {\exp \left( \frac {h\nu}{k_B T_{BB}}\right) - 1} d\nu} {C_n  \int_{0}^{\infty} \nu^2 \frac {1} {\exp \left( \frac {h\nu}{k_B T_{BB}}\right) - 1} d\nu}\\
& \langle \epsilon_{av} \rangle \approx 2.701 \text{ } k_B \text{ } T_{BB}
\end{aligned}
\end{equation}
Here, $\nu$ denotes the radiation frequency. Considering the total energy $E$ , the radiation pressure\cite{49} $P = E/3V$ (where $V$ is the volume), and the average number of particles $\langle n_{av} \rangle$, the equation may be rearranged as
\begin{equation}
PV \sim 0.9 \text{ } \langle n_{av} \rangle \text{ } k_B \text{ } T_{BB}.
\end{equation}
For the estimation of free energy and work potential, we require one additional aspect from the quantum mechanical description of light–matter interaction. In a matter–photon system undergoing a unitary process of absorption or emission, the entropies of the subsystems ($A$ and $B$, with entropies $S_A$ and $S_B$) are related to the total entropy ($S_{AB}$) through the Araki–Lieb inequality \cite{52,58,68} as follows:
\begin{equation}
|S_A - S_B| \le S_{AB} \le S_A + S_B
\end{equation}
On the other hand, the total entropy ($S_{AB}$) can also be expressed as\cite{52}
\begin{equation}
 S_{AB} = S_A + S_B - I(A:B)
\end{equation}
where the second term ($I(A:B)$) represents the entropy associated with subsystem entanglement. In a typical photon-absorption process, the interaction between radiation and matter (e.g., in an energy-harvesting material) occurs primarily through the electronic wavefunction of the material. Therefore, without resorting to many-body physics, we assume that absorption is a local process occurring predominantly via single-photon events (unless otherwise stated), and that the interaction between the radiation field and the absorber can be treated from an electronic perspective. For thermal radiation, such as that from the Sun, the initial state of the radiation field is a mixed state with entropy $S_A(0)$, whereas the electronic ground state is a pure state, $S_B(0) = 0$. Assuming no correlation between the subsystems, the overall entropy can be written as $S_{AB} = S_A(0)$. Consequently, under a unitary evolution that conserves total entropy, the following inequality can be expressed for the time evolution of subsystem entropies using the above-mentioned expression:
\begin{equation}
 I(A:B) = S_B (t) + S_A (t) - S_A (0)
\end{equation}
Imposing the non-negativity condition $I(A:B) \ge 0$, the following expression can be written for the present case:
\begin{equation}
 \Delta S_B (t) \ge - \Delta S_A (t)  \text {  where, }  \Delta S_X = \left( S_X (t) -  S_X (0) \right)
\end{equation}
We may further use the following approximation: in many of these types of open-system processes, the electron–photon interaction can be treated in the weak-coupling, Born–Markovian limit. Under this assumption, electron–photon field correlations (discussed in the previous section) can be treated as negligible. This approximation gives $\Delta S_B(t) \approx -\Delta S_A(t)$. Therefore, by monitoring the entropy change of one subsystem ($B$), one can infer the entropy change of the other subsystem ($A$).\\
For ideal systems, such as isolated atoms, photon absorption is expected to leave the system in a pure excited state. In contrast, in real materials such as solids, entropy changes may arise from the formation of electron–hole pairs (excitons) \cite{79} and from the multiplicity of accessible configurations, including possible spin arrangements \cite{79}. In most solid-state materials composed of light elements, which exhibit low spin–orbit coupling and are commonly used for energy-harvesting applications, the transient excited state is most often found in a singlet configuration \cite{79}. Although a singlet state is overall a pure state, the reduced entropy \cite{52,58} of a subsystem, such as the electron, can be estimated as $k_B \ln 2$ when the two particles are maximally entangled. Even in the presence of a small admixture of triplet states, which can occur in materials containing heavy elements, the subsystem entropy remains close to $k_B \ln 2$ per particle. Accordingly, we take $k_B \ln 2$ as the transient entropy change of the matter (electron) associated with the absorption of a single photon. We then apply the inequality described in the previous section to the unitary evolution of the electron–photon system, concluding that the absorption of a photon by an electron leaves the photon field with an entropy decrease of $\sim k_B \ln 2$ for a single-photon absorption.
With this information, we can now estimate the loss of work potential by evaluating the free energy \cite{24,25} associated with photon absorption. To quantify the loss of work potential during the absorption process, we first note that unitary evolution conserves the total entropy of the universe. Consequently, in a scenario where all subsystems are at thermal equilibrium and share the same temperature, the concepts of free energy and exergy become less straightforward to interpret. However, in the present case, the field and the sample are at different temperatures and exchange energy during the absorption process. This nonequilibrium situation allows for a meaningful evaluation of the loss of work potential. Accordingly, the free-energy\cite{24,25} changes of the photons (subscript with $p$) and the sample (subscript with $s$) during absorption can be written as follows:
\begin{equation}
\delta F_p = - \delta U + T_p |\delta S_a| \text { and } \delta F_s =  \delta U - T_s |\delta S_a|
\end{equation}
Therefore, the total free energy change can be written as follows,
\begin{equation}
\begin{aligned}
\delta F_t = \delta F_p + \delta F_s &= |\delta S_a| \left( T_p - T_s\right)\\
                        &= \left( 1 - \frac{T_s}{T_p}\right) T_p |\delta S_a|\\
                        &= \eta_c T_p |\delta S_a|
\end{aligned}
\end{equation}
where $\eta_c$ is related to the Carnot efficiency \cite{24,25}. We identify this total free energy increase as the loss of work potential in the absorption process. This loss captures, to some extent, the effects of the microscopic details associated with the energy transfer process from the emitter (here, the Sun) to the absorber (here, the sample). Considering $T_p$ ($\sim$ 5778 kelvin) as the temperature of the Sun and $T_s$ ($\sim$ 300 kelvin) as the sample temperature, we may estimate the available work potential as:
\begin{equation}
\begin{aligned}
\text {Work potential($\mathcal{W}_f$)} &= \delta U - \eta_c T_{sun} |\delta S_a|\\
                      &= E - 0.948 \times  \langle n_{av} \rangle k_B T_{sun} \ln 2\\
                     &= E - \frac {0.948 \times \ln 2}{2.701}E\\
\text{Work potential($\mathcal{W}_f$)/E} & \sim 75.67\text{$\%$}
\end{aligned}
\end{equation}
Therefore, within this framework, approximately 75.67$\%$ of the absorbed photon energy can be considered usable. Note that we have used the term work potential ($\mathcal{W}_f$) to indicate that it is derived from the free energy. However, for a more accurate calculation, one may replace the Carnot efficiency term $\eta_c$ with the R. Petela exergy expression (\textbf{Equation 8}). This yields an efficiency of approximately 76.11$\%$ for single-photon processes. One may expand (\textbf{Equation 16}) by using (\textbf{Equations 4-6}), considering $T_p$ as the emitter (Sun) temperature and the associated entropy change during emission as $\delta S_e$. With this, \textbf{Equation 16}, relating to the Petela equation, becomes
\begin{equation}
\begin{aligned}
\text {Work potential($\mathcal{W}_r$)} & = E_{\text{ex}}^{p} - \eta_c dE \frac {|\delta S_a|}{|\delta S_e|}\\
\text {Work potential($\mathcal{W}_r$)} & = E_{\text{ex}}^{p} \left(1 - \frac {|\delta S_a|}{|\delta S_e|}\right)
\end{aligned}
\end{equation}
\textbf{Equation 18} provides a more general expression for work potential estimates for solar radiation. This includes the R. Petela exergy expression ($E_{\text{ex}}^{p}$) as well as the entropy exchange during photon emission and absorption. The estimated value of the work potential $\mathcal{W}_r$ using \textbf{Equation 18} is around 75.2$\%$. Note that in the derivation by R. Petela, the microscopic detail associated with the energy transfer path is not considered. Consequently, the resulting expression represents the maximum work potential under ideal conditions. However, the actual energy transfer path (here, from the Sun to the sample) may involve additional sources of loss of work potential. Accounting for these microscopic details, particularly those associated with the absorption process, leads to the expression given in \textbf{Equation 18}. It is understandable from \textbf{Equation 18} that, if one can reduce $\delta S_a$, one may achieve a higher work potential. A straightforward way to achieve lower $\delta S_a$ is to use a two-photon absorption process. In a second-order two-photon process, the system absorbs two photons through a second-order transition, generating a single electron–hole pair; thus, twice the photon energy is absorbed for the same entropy cost of $k_B \ln 2$. This lower entropy cost manifests as higher efficiency, and, following an analogous derivation, the work potential is calculated to be approximately 84$\%$ for a two-photon absorption process. Based on these estimates, we consider the maximum thermodynamic work potential for solar energy conversion under standard conditions to be approximately 76$\%$. For second-order two-photon processes, the efficiency increases to approximately 84$\%$. While a rigorous treatment would require a detailed description of the photon–electron interaction, the underlying unitary evolution of the coupled electron–photon field, and additional factors such as the differing probabilities of first- and second-order absorption processes, these estimates provide a practical measure of the work availability of radiation. We validate this measure by calculating the solar-to-usable-energy conversion efficiency of a solar cell under direct sunlight, as presented below.
\subsection {Limitations of solar energy conversion}
\begin{figure}[t]
\begin{center}
\includegraphics[width=1.0\columnwidth]{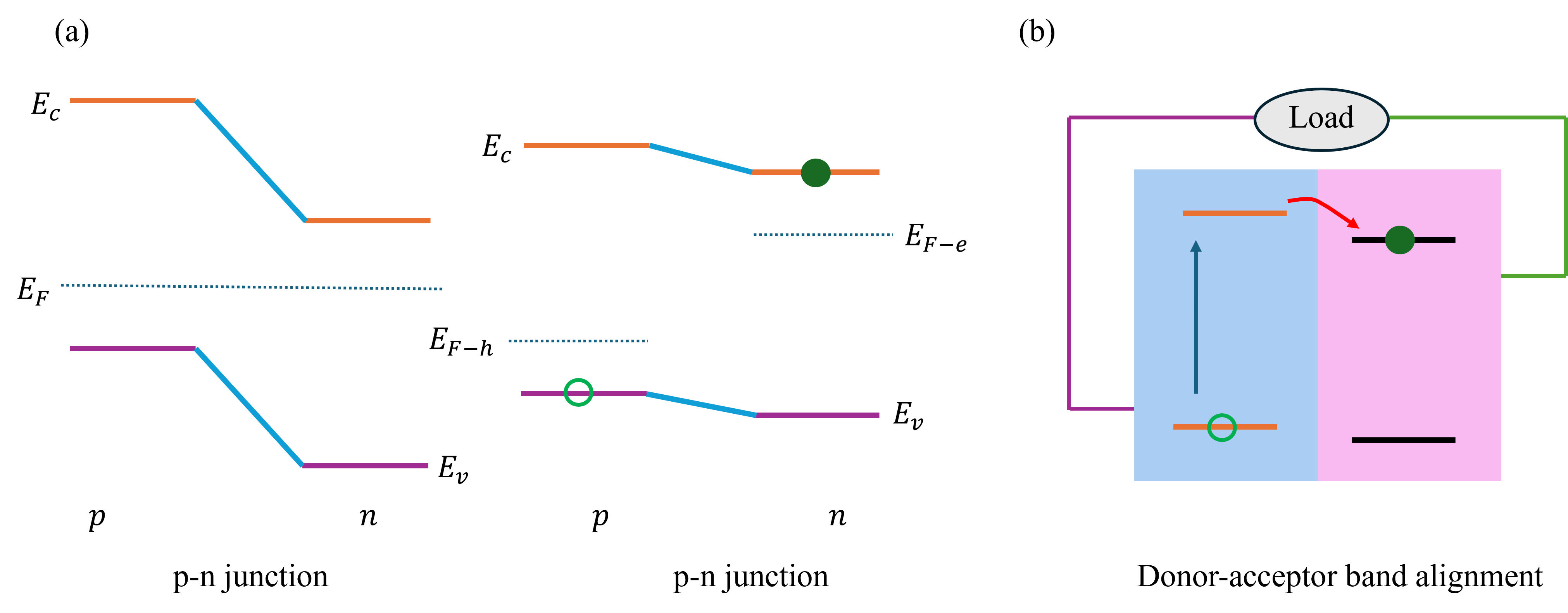}
\caption{Panels (a) and (b) illustrate typical methods used for electron–hole separation in solar-cell devices: formation of a p-n junction (a) and donor–acceptor band alignment (b).}
\label{fig2}
\end{center}
\end{figure}
In order to validate the usability of the free energy estimate and to develop a simplistic model for evaluating the S–Q limit \cite{44,45} under various circumstances, we employ the simplified model presented below to investigate energy-harvesting efficiencies in detail. The following model allows for the inclusion of the effects of spontaneous emission (electron-hole recombination after excitation, not to be confused with thermal blackbody radiation from the sample) on solar efficiencies, which are not explicitly considered in the limitations estimated by S-Q. A typical description of a solar energy harvesting mechanism is provided in \textbf{Figures 2a and 2b}. Once a material with a band gap $E_g$ absorbs a photon (energy $E_{ex} = E_g = h\nu$, where $\nu$ is the frequency of the light), the electron and the hole generally form a bound electron-hole pair known as an exciton\cite{79,80,81,82}, with binding energy referred to as the exciton binding energy ($E_B$)\cite{80,81,82,83}. Due to the formation of this bound state, the electron-hole pair has slightly less energy than the band gap ($E_g$) of the material. In other words, one may define an optical band gap\cite{83} as $E_g - E_B$.\\
To harvest energy from an exciton, the usual approach is to spatially separate the electron–hole pair over a larger distance, thereby suppressing processes such as spontaneous emission and enabling the charges to be collected in an external circuit. In most solar cells, this is achieved via band alignment between dissimilar materials\cite{79,84} or through the formation of a p–n junction \cite{85,86} (see \textbf{Figure 2}).
The separation of charge carriers over a larger distance can occur via either an energy-lowering (band-alignment) mechanism or an entropy-driven mechanism that reduces the Coulomb attraction between the electron and the hole \cite{79,84}. For an energy-lowering process, such as band alignment, the magnitude of the loss may be approximated by the exciton binding energy. However, if the process is entropy-driven, the loss associated with spontaneous electron–hole separation over a larger distance is difficult to estimate. Nonetheless, we may crudely approximate it by setting the effective energy to be $(E_g - E_B)$, similar to the energy-driven process. Once the electron-hole pair is created, a voltage-often referred to as the open-circuit voltage\cite{85,86,87,88,89}-is generated.
The maximum voltage magnitude is expected to be proportional to the bandgap ($E_g$) of the material \cite{85,86,87,88} and, therefore, to the energy ($h\nu$) of the absorbed photon. To express the maximum solar efficiency, we neglect the reduction of the open-circuit voltage through other mechanisms (e.g., disorder, tail states, electrode contacts, etc.)\cite{85,86,87,90}. We express the distribution of the variable that quantifies the electron-hole pairs generated to flow through the external circuit as exponential in energy ($h\nu$). Therefore, within the framework of cumulative distribution theory \cite{91}, the cumulative distribution function of this exponential distribution can be written as $\mathcal{Q} \propto 1 - \left[\exp \left(-K^\prime h \nu /K^{\prime\prime}\right)\right]$, where $K^{\prime\prime}$ is a constant associated with the mean of the exponential distribution is later found to relate to an effective energy barrier. $K^\prime$ is a proportionality constant, indicating that the maximum open-circuit voltage scales with the band gap of the material. From here on, we denote $\mathcal{Q}$ as the probability that an excited electron-hole pair can be collected in an external circuit.
Considering the optical gap ($E_g - E_B$) as the `effective' energy (as previously discussed), the probability becomes $\mathcal{Q} \propto 1 - \left[\exp\left\{-K^\prime \left( h \nu - E_B\right)/K^{\prime\prime}\right\}\right]$. For most materials, the exciton binding energy is inversely proportional to the dielectric constant of the material\cite{80,81}, which decreases nearly exponentially as the bandgap increases\cite{92,93}. Therefore, we can approximate $E_B$ as $ E_B \propto K_2 \exp \left( E_g \right)$. An approximately linear variation\cite{83} may also be used for certain materials, such as organic semiconductors. We set the value of $K_2$ to be approximately $0.001$ based on the known exciton binding energy of $\sim 0.5$ eV for $E_g \approx 6.2$ eV\cite{82}. Note that this dependence of
$E_b$ on the band gap is approximately valid for band-gap values between $\sim$ 0.001 eV and $\sim$ 6.2 eV, which are used in all the calculations. Therefore, we can write the probability (noting that $E_g \approx h\nu$ ) as $\mathcal{Q} \propto 1 - \left[\exp\left\{-K^\prime \left( h \nu - 0.001 \exp \left(h \nu \right)\right)/K^{\prime\prime} \right\}\right]$ . Finally, at any stage of electron-hole separation, other recombination processes, such as spontaneous emission (radiative recombination of the electron-hole pair, not to be confused with thermal blackbody radiation) or nonradiative thermal recombination, will decrease the probability that the electron-hole pair is collected in the circuit. For simplicity, we assume that nonradiative recombination across the bandgap is minimal in this process (e.g., disorder and trap-state densities are low), whereas spontaneous emission cannot be completely neglected. The probability of spontaneous emission can be described using the Einstein coefficients\cite{6,9,48}. We may express the spontaneous emission transition probability (rate) as being proportional to the frequency,$\propto \nu^3/c^3$, where $c$ is the speed of light in free space\cite{6,9,48}. Including this term in the above probability gives the probability that an electron-hole pair created by the absorption of a photon of energy ($ h\nu \approx E_g $) is collected in the circuit as
\begin{equation}
\begin{aligned}
\text{$\mathcal{Q}$} &= \left[1 - \exp \left(-M/K_1 \right)\right] \text{for} M \ge 0\\
\text{$\mathcal{Q}$} &= 0 \text{ for } M < 0\\
\text{Where, } M &= \left\{ h \nu - 0.001 \exp \left(h \nu \right) - K_3 \left( \nu^3/c^3 \right) \right\}
\end{aligned}
\end{equation}
Within the present framework, the parameter $K_1$ is interpreted as an effective energy barrier that may incorporate multiple physical effects. Throughout most of the calculations, its value is fixed at approximately 250 meV. Subsequently, we examine the dependence of the maximum efficiency on the magnitude of this effective barrier. The constant $K_3$ is used for parameterization as well as to account for the dependence of the transition probability of spontaneous emission on other effects, such as coupling strength.  Hereafter, the term $\mathcal{Q}$ is referred to as the ``\emph{quality factor}'', which quantifies how closely a sample approaches the ideal value of $\mathcal{Q}=1$. To calculate the fraction of photons that can be converted into electricity, we use the following method. First, we assume that the number of photons reaching Earth from the Sun can be described by a Planck distribution, $N_t \left( h\nu \right)$\cite{6,44}. The material absorbs all photons ($N_{abs}$) with energies above its band gap. However, after excitation with energy higher than the band gap, the material loses some energy due to phonon scattering (nonradiative heat loss) and relaxes to the band-gap energy. In other words, all processes occurring in the material after excitation take place at the band-gap energy. Therefore, we may write the total number of photons incident on the sample and the number of photons absorbed as\cite{6,44}
\begin{equation}
N_t \propto \int_{\nu_{min}}^{\nu_{max}} \nu^2 \frac {1} {\exp \left( \frac {h\nu}{k_B T_{sun}}\right) - 1} d\nu
\end{equation}
\begin{equation}
N_{abs} \propto \int_{E_g}^{\nu_{max}} \nu^2 \frac {1} {\exp \left( \frac {h\nu}{k_B T_{sun}}\right) - 1} d\nu
\end{equation}
Here, $T_{sun}$ denotes the approximate surface temperature of the Sun, and the values of $\nu_{min}$ and $\nu_{max}$ are set to 0.001 eV and 6.2 eV, respectively, for all calculations. The calculations were performed using typical numerical methods, including numerical integration techniques. To evaluate the energy/power conversion efficiency, we estimate the solar energy reaching Earth using a Planck distribution\cite{6,44}, given by
\begin{equation}
E_t \propto \int_{\nu_{min}}^{\nu_{min}} \nu^3 \frac {1} {\exp \left( \frac {h\nu}{k_B T_{sun}}\right) - 1} d\nu
\end{equation}
The number of photons absorbed by the material remains the same as given by \textbf{Equation 21}. With this, the energy conversion efficiency can be expressed as
\begin{equation}
\begin{aligned}
\eta_E &=  0.76 \text{ } \mathcal{Q} \frac {N_{abs}\left(E_g - E_b \right)}{h E_t}\\
       &= 0.76 \text{ } \mathcal{Q} \frac {N_{abs}\left(h\nu - 0.001 \exp \left(h\nu\right) \right)}{h E_t}\\
\text {Therefore,  } \eta_E &= 0.76 \times \mathcal{Q} \times \Lambda_{\mathrm{L}}
\end{aligned}
\end{equation}
\begin{figure*}[h!t]
\begin{center}
\includegraphics[width=1.0\columnwidth]{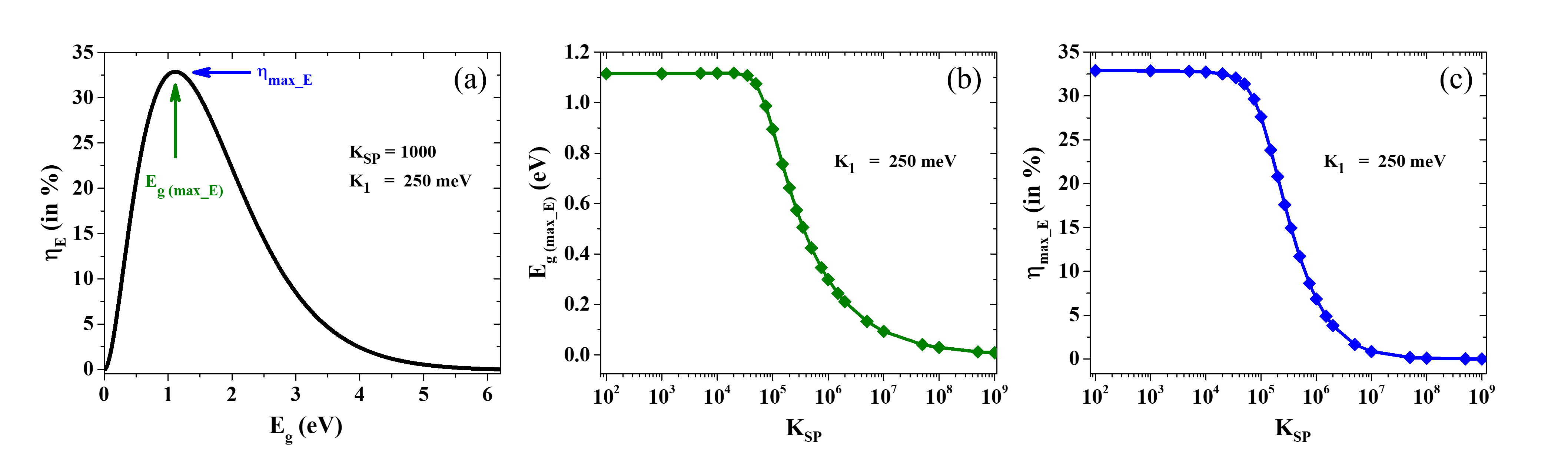}
\caption{Panel (a) shows the variation of the energy (power) conversion efficiency, $\eta_E$, as a function of the material bandgap, $E_g$, at fixed values of $K_{sp} = 1000$ and $K_1 = 250$ meV. The maxima, corresponding to $E_{g(max\_E)}$ and $\eta_{max\_E}$ at fixed values of $K_{sp} = 1000$ and $K_1 = 250$ meV, are indicated by the blue and olive lines, respectively. Panels (b) and (c) show the variation of $E_{g(max\_E)}$ and $\eta_{max\_E}$ for different values of $K_{sp}$, at a fixed value of $K_1 = 250$ meV. Note that $K_{sp}$ refers here only to a proportionality constant related to the electron–hole lifetime and does not give the exact value of the electron–hole lifetime.}
\label{fig3}
\end{center}
\end{figure*}
\begin{figure*}[t]
\begin{center}
\includegraphics[width=1.0\columnwidth]{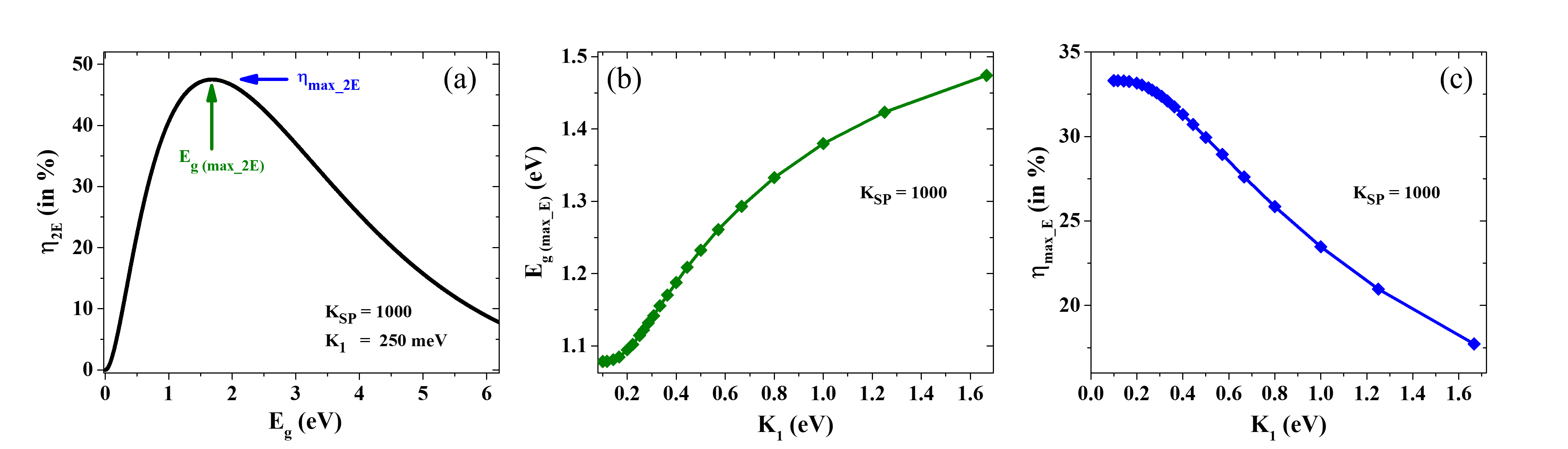}
\caption{Panel (a) shows the variation of the energy (power) conversion efficiency, $\eta_{2E}$, as a function of the material bandgap, $E_{g}$, at fixed values of $K_{sp} = 1000$ and $K_1 = 250$ meV, considering the two-photon absorption processes (see manuscript for details). Panels (b) and (c) show the variation of $E_{g(max\_E)}$ and $\eta_{max\_E}$, respectively, as a function of $K_1$ at a fixed value of $K_{sp} = 1000$, for the one photon absorption study.}
\label{fig4}
\end{center}
\end{figure*}
Note that the multiplication by 0.76 is included here to account for the maximum work availability discussed in the previous section. Note that we have used the symbol $\Lambda_{\mathrm{L}}$ to identify the effect of the ``\emph{loss-modified retention factor}''. With this, the overall efficiency becomes $\eta_E = 0.76 \times \mathcal{Q} \times \Lambda_{\mathrm{L}}$. Using the above equation, we estimate the energy conversion efficiency, $\eta_E$, as a function of the material’s band gap ($E_g$) for a fixed $K_{sp} = K_3 /c $ value of 1000, as shown in \textbf{Figure 3a}. As the band gap increases, $\eta_E$ increases and, after reaching a maximum at a band gap of approximately 1.11 eV, begins to decrease continuously with further increases in the band gap. We identify this maximum as the most suitable band gap, $E_{g(max\_E)}$, corresponding to the maximum photon conversion efficiency, $\eta_{max\_E}$, for a $K_{sp}$ value of 1000. \textbf{Figures 3b and 3c} show the variation of $\eta_{max\_E}$ and $E_{g(max\_E)}$ with changes in $K_{sp}$ over a wide range, from approximately $10^9$ to $10^2$. A wide variation in spontaneous emission rates has indeed been observed in various samples under different conditions\cite{94,95}, for example, by increasing the spatial separation of the electron and hole, which reduces their wavefunction overlap and consequently the emission rate. We find that for samples with a very high probability of spontaneous emission (large $K_{sp}$), the photon conversion efficiency ($\eta_{max\_E}$) decreases substantially. As the spontaneous emission probability decreases, the photon conversion efficiency increases and reaches a maximum value of approximately 32.9$\%$. Thereafter, $\eta_{max\_E}$ remains nearly independent of further reductions in $K_{sp}$. The band gap at which $\eta_E$ reaches its maximum, $E_{g(max\_E)}$, also varies with $K_{sp}$ and approaches a value of approximately 1.11 eV at lower $K_{sp}$ values. These values are very close to those obtained for solar energy conversion efficiency using various other methods\cite{27,44}. Note that the above calculations are not restricted to a single material (or a single p-n junction) but can be easily extended to multi-junction solar cells.\\

We calculate the energy conversion efficiency for a trilayer solar cell composed of materials with three different band gaps, $E_{g\_1} \sim$ 1.11 eV, $E_{g\_2} \sim$ 0.8 eV, and $E_{g\_3} \sim$ 0.5 eV, using the following equation:
\begin{equation}
\begin{aligned}
\eta_{E\_tri} & = 0.76 \times \sum\limits_{j} \left[\mathcal{Q}_j \times \Lambda_{\mathrm{L}\_j}\right] = 0.76 \times \frac {\sum\limits_{j} \mathcal{Q}_j \left( E_{g\_j} - E_{B\_j} \right) N_{abs\_j}}{h E_t}\\
&\text {Where,}\\
N_{abs\_j} & = \int_{E_{g\_j}}^{\nu_{max\_j}}\nu^2 \frac {1} {\exp \left( \frac {h\nu}{k_B T_{sun}}\right) - 1} d\nu
\end{aligned}
\end{equation}
Here, $Q_j$ and $E_{B\_j}$ represent the probability of electron extraction and the exciton binding energy of the $j^{th}$ material (or junction), respectively. Note that $\nu_{max\_j}$ will be different for the three materials. For the material with the highest band gap ($E_{g\_1} \sim$ 1.11 eV), $\nu_{max\_1}$ is 6.2 eV; however, for the lower band gap material (e.g., $E_{g\_2} \sim$ 0.8 eV), $\nu_{max\_2}$ is approximately 1.11 eV, and so on. These calculations indicate that the maximum energy conversion efficiency, with the typical band gaps of the trilayer junction chosen here, is approximately 44.7$\%$ for $K_{sp} = 1000$ and $K_1 = 250$ meV. This value also closely matches the maximum energy conversion efficiency obtained using a multilayer (trilayer) p-n junction\cite{46,47}.\\

Finally, we aim to provide a theoretical estimate of the maximum efficiency for an upconverted solar cell, in which some of the photons below the band-gap energy are absorbed by the material through nonlinear two-photon absorption\cite{96}. For this purpose, we assume that all photons above the band gap are absorbed in the normal way, whereas photons with energies below the band gap but above half the band gap are absorbed via two-photon processes. In this case, the conversion efficiency can be written as:
\begin{equation}
\begin{aligned}
&\eta_{2E} = \eta_{1ph} +  \eta_{2ph}\\
&\eta_{1ph} = \frac {0.76 \times \mathcal{Q} \left\{ N_{abs}\left(E_g - E_b \right)\right\}} {h E_t}\\
&\eta_{2ph} = \frac {0.84 \times 0.5 \times \mathcal{Q} \left\{ N_{abs2}\left(E_g - E_b \right)\right\}}{h E_t}\\
&\text{Where, } N_{abs2} \propto \int_{0.5E_g}^{E_g} \nu^2 \frac {1} {\exp \left( \frac {h\nu}{k_B T_{sun}}\right) - 1} d\nu
\end{aligned}
\end{equation}
The factor 0.5 arises because two photons from the spectrum combine to create a single electron-hole pair. \textbf{Figure 4a} shows the variation of the estimated efficiency, $\eta_{2E}$, as a function of the band gap of the material. The variation of the efficiency is very similar to that of the single-photon absorption case. However, for the present scenario, a substantial increase in the maximum efficiency is observed. The band gap at which the maximum efficiency occurs has shifted to higher values. In this case, the maximum efficiency reaches approximately 48$\%$, significantly higher than the value predicted by the S-Q model. The most suitable band gap for the present scenario is around 1.68 eV.\\

Note that the effect of the effective energy barrier $K_1$ has not been discussed in the calculations presented thus far. \textbf{Figures 4b and 4c} show the variation of $\eta_{max\_E}$ and $E_{g(max\_E)}$ with $K_1$, calculated for a fixed low value of $K_{sp}=1000$. We find that as $K_1$, or equivalently the effective barrier, decreases, the efficiency increases and eventually saturates at approximately 33.3$\%$ for very low values of $K_1$. Simultaneously, the optimum band gap decreases and approaches a saturation value of approximately 1.1 eV. For the trilayer sample, the saturation efficiency at low $K_1$ reaches nearly 46$\%$. Based on these trends, we choose $K_1 = 250$ meV, which corresponds approximately to the onset of efficiency saturation and is therefore used in all preceding calculations. From the work potential perspective, one would expect the maximum saturation to be 76$\%$; however, we find that it instead saturates at around 33.3$\%$. This limitation arises from the fact that below-bandgap radiation is not usable in single-photon absorption processes, and from the approximation that the process occurs at the bandgap level: any electron excited with energy greater than the bandgap loses its excess energy as heat and relaxes to the band edge. This thermalization loss further limits the energy conversion efficiency to approximately 33.3$\%$. Therefore, we suggest that if such thermal losses can be reduced, for example through multilayer formation or other methods, efficiencies higher than the S–Q limit may be achieved, although the maximum efficiency may ultimately saturate around 76$\%$. Finally, note that in our calculations, the sample temperature appears through the temperature dependence of the bandgap $E_g$\cite{97}. We believe that additional temperature dependence may arise from the $K_1$ parameter, if its explicit form is known, in the probability function (\textbf{Equation 19}), which we have set to 250 meV for most calculations.

\section{Conclusion}

In conclusion, given current limitations in understanding the thermodynamics of light and light–matter interactions, we employ a simple, approximate approach to examine the thermodynamic behavior of light and evaluate its exergy and work potential under different conditions, such as single-photon and two-photon absorption. Specifically, this approach enables the inclusion of microscopic details of the energy transfer process, thereby providing a more reliable estimate of the exergy of radiation. Within this simplified framework, we suggest that the thermodynamic limit of light-to-usable-energy conversion efficiency is approximately 76$\%$, which is significantly higher than the S–Q limit. To validate this thermodynamic estimate, we provide a simplified model and calculate the solar-to-usable-energy conversion efficiencies under various conditions (single-junction and multi-junction), including effects such as spontaneous emission. The commonly observed efficiency of $\sim$ 33.3$\%$, known as the S–Q limit, was effectively reproduced using this model and the exergy and work potential estimates, supporting the validity of both the exergy calculations and the simplified model. The presented model and the exergy and work potential estimates may also help explain the observation\cite{91} that the open-circuit voltage obtained in most solar cells is $\sim$ 25$\%$ lower than the theoretical maximum. Furthermore, the model allowed us to evaluate the role of spontaneous emission in limiting efficiency. We observed that a high spontaneous emission rate reduces energy conversion efficiency; however, this effect can be partially mitigated through appropriate band alignment or the formation of a p–n junction. On the other hand, thermal loss of electrons—specifically after excitation with energy higher than the bandgap—is a major factor limiting solar energy conversion efficiency. If such losses can be reduced, for example through the use of multilayered junctions, higher light-to-usable-energy conversion efficiencies can be achieved. Furthermore, upconversion two-photon solar cells are particularly promising, as they possess high exergy content and a theoretical efficiency approaching 48$\%$, well above the S–Q limit. Considering these findings, we suggest that the commonly reported or estimated efficiencies of approximately 33.3$\%$ to 48$\%$ under different conditions should be regarded as conditional efficiencies. In contrast, the true thermodynamic efficiency of light-to-usable-energy conversion is approximately 76$\%$ for single-photon absorption processes. However, a precise estimation of this efficiency requires a more detailed understanding of light–matter interactions and the thermodynamics of light.

\section{Acknoledgement}

The author thanks the Indian Institute of Science, Bangalore, for providing the facilities to conduct this work. Parts of the manuscript was proofread for grammatical accuracy using ChatGPT, an AI-based language tool.

\section{Data availability}

The data that support the findings of this study are available from the corresponding author upon reasonable request.




\begin{thebibliography}{}

\bibitem{1} D. J. Griffiths, and D. F. Schroeter, \emph{Introduction to Quantum Mechanics} (Cambridge University Press, 3rd edition, 2018).

\bibitem{2} P. A. M. Dirac, The quantum theory of the emission and absorption of radiation, Proc. R. Soc. Lond., Ser. A, \textbf{114}, 243 (1927). 10.1098/rspa.1927.0039

\bibitem{3} P. W. Milonni, Why spontaneous emission?, Am. J. Phys., \textbf{52}, 340 (1984). 10.1119/1.13886

\bibitem{4} B. Romeira, and A. Fiore, Purcell Effect in the Stimulated and Spontaneous Emission Rates of Nanoscale Semiconductor Lasers, IEEE J. Quantum Electron., \textbf{54}, 1 (2018). 10.1109/JQE.2018.2802464

\bibitem{5} J. Gea-Banacloche, M. O. Scully, and M. S. Zubairy, Vacuum Fluctuations and Spontaneous Emission in Quantum Optics, Phys. Scr., \textbf{T21}, 81 (1988). 10.1088/0031-8949/1988/T21/015

\bibitem{6} A. Beiser, \emph{Concepts of Modern Physics} (McGraw-Hill, sixth edition, 2003).

\bibitem{7} D. S. Lemons, W. R. Shanahan, L. Buchholtz, and M. Gyeviki, \emph{On the Trail of Blackbody Radiation: Max Planck and the Physics of His Era} (The MIT Press, 2022).

\bibitem{8} C. Johnson, \emph{Mathematical Physics of BlackBody Radiation} (Icarus iDucation, 2012).

\bibitem{9} S. Weinberg, \emph{The Quantum Theory of Fields} (Cambridge University Press, 2005).

\bibitem{10} M. O. Scully, and M. S. Zubairy, \emph{Quantum Optics} (Cambridge University Press, 1997).

\bibitem{11} H. Carmichael, \emph{An Open Systems Approach to Quantum Optics} (Springer-Verlag, 1993).

\bibitem{12} H. Khosravi, and R. Loudon, Vacuum field fluctuations and spontaneous emission in the vicinity of a dielectric surface, Proc. R. Soc. Lond. A Math. Phys. Sci., \textbf{433}, 337 (1991). 10.1098/rspa.1991.0052

\bibitem{13} M. A. Weinstein, Thermodynamics of Radiative Emission Processes, Phys. Rev., \textbf{119}, 499 (1960). 10.1103/PhysRev.119.499

\bibitem{14} O. Kafri, Entropy and Temperature of Electromagnetic Radiation, Nat. Sci., \textbf{11}, 323 (2019). 10.4236/ns.2019.1112035

\bibitem{15} A. Rueda, On the irreversible thermophysics of radiative processes, Found. Phys., \textbf{4}, 215 (1974). 10.1007/BF00712688

\bibitem{16} D. Mauzerall, Thermodynamics of primary photosynthesis, Photosynth. Res., \textbf{116}, 363 (2013). 10.1007/s11120-013-9919-x

\bibitem{17} P. T. Landsberg, and G. Tonge, Thermodynamic energy conversion efficiencies, J. Appl. Phys., \textbf{51}, R1 (1980). 10.1063/1.328187

\bibitem{18} S. M. Jeter, Maximum conversion efficiency for the utilization of direct solar radiation, Sol. Energy, \textbf{26}, 231 (1981). 10.1016/0038-092X(81)90207-3

\bibitem{19} A. Delgado-Bonal, Entropy of radiation: the unseen side of light, Sci. Rep., \textbf{7}, 1642 (2017). 10.1038/s41598-017-01622-6

\bibitem{20} A. Rueda, Entropy production for a near-equilibrium isolated system of thermalizing matter and radiation, J. Chem. Phys., \textbf{58}, 3320 (1973). 10.1063/1.1679658

\bibitem{21} A. D. Kirwan, Intrinsic photon entropy? The darkside of light, Int. J. Eng. Sci., \textbf{42}, 725 (2004). 10.1016/j.ijengsci.2003.09.005

\bibitem{22} A. Ore, Entropy of Radiation, Phys. Rev., \textbf{98}, 887 (1955). 10.1103/PhysRev.98.887

\bibitem{23} P. Rosen, Entropy of Radiation, Phys. Rev., \textbf{96}, 555 (1954). 10.1103/PhysRev.96.555

\bibitem{24} P. W. Atkins, and J. De Paula, \emph{Atkins’ Physical Chemistry} (W. H. Freeman, 8th edition, 2006).

\bibitem{25} M. J. Moran, and H. N. Shapiro, \emph{Fundamentals of Engineering Thermodynamics}, (John Wiley $\&$ Sons, 5th edition, 2006).

\bibitem{26} E. Penocchio, R. Rao, and M. Esposito, Nonequilibrium thermodynamics of light-induced reactions, J. Chem. Phys., \textbf{155}, 114101 (2021). 10.1063/5.0060774

\bibitem{27} A. De Vos, and H. Pauwels, On the thermodynamic limit of photovoltaic energy conversion, Appl. Phys., \textbf{25}, 119 (1981). 10.1007/BF00901283

\bibitem{28} D. Juretic, Efficiency of free energy transfer and entropy production in photosynthetic systems, J. Theor. Biol., \textbf{106}, 315 (1984). 10.1016/0022-5193(84)90033-X

\bibitem{29} D. Juretic, and P. Županovic, Photosynthetic models with maximum entropy production in irreversible charge transfer steps, Comput. Biol. Chem., \textbf{27}, 541 (2003). 10.1016/j.compbiolchem.2003.09.001

\bibitem{30} C. D. Andriesse, and M. J. Hollestelle, Minimum entropy production in photosynthesis, Biophys. Chem., \textbf{90}, 249 (2001). 10.1016/S0301-4622(01)00146-6

\bibitem{31} R. C. Jennings, E. Engelmann, F. Garlaschi, A. P. Casazza, and G. Zucchelli, Photosynthesis and negative entropy production, BBA Bioenergetics, \textbf{1709}, 251 (2005). 10.1016/j.bbabio.2005.08.004

\bibitem{32} X. L. Ruan, S. C. Rand, and M. Kaviany, Entropy and efficiency in laser cooling of solids, Phys. Rev. B, \textbf{75}, 214304 (2007). 10.1103/PhysRevB.75.214304

\bibitem{33} C. E. Mungan, Thermodynamics of radiation-balanced lasing, J. Opt. Soc. Am. B, \textbf{20}, 1075 (2003). 10.1364/JOSAB.20.001075

\bibitem{34} C. E. Mungan, Radiation thermodynamics with applications to lasing and fluorescent cooling, Am. J. Phys., \textbf{73}, 315 (2005). 10.1119/1.1842732

\bibitem{35} M. O. Scully, Laser entropy: from lasers and masers to Bose condensates and black holes, Phys. Scr., \textbf{95}, 024002 (2020). 10.1088/1402-4896/ab41fc

\bibitem{36} A. Cuadras, J. Yao, and M. Quilez, Determination of LEDs degradation with entropy generation rate, J. Appl. Phys., \textbf{122}, 145702 (2017). 10.1063/1.4996629

\bibitem{37} J. Xue, Z. Li, and R. J. Ram, Irreversible Thermodynamic Bound for the Efficiency of Light-Emitting Diodes, Phys. Rev. Appl., \textbf{8}, 014017 (2017). 10.1103/PhysRevApplied.8.014017

\bibitem{38} D. S. Lebedev, and L. B. Levitin, Information transmission by electromagnetic field, Inf. Control., \textbf{9}, 1 (1966). 10.1016/S0019-9958(66)90074-X

\bibitem{39} S. E. Harris, Electromagnetically induced transparency and quantum heat engines, Phys. Rev. A, \textbf{94}, 053859 (2016). 10.1103/PhysRevA.94.053859

\bibitem{40} S. J. van Enk, and G. Nienhuis, Entropy production and kinetic effects of light, Phys. Rev. A, \textbf{46}, 1438 (1992). 10.1103/PhysRevA.46.1438

\bibitem{41} M. Santillán, G. A. de Parga, and F. Angulo-Brown, Black-body radiation and the maximum entropy production regime, Eur. J. Phys., \textbf{19}, 361 (1998). 10.1088/0143-0807/19/4/008

\bibitem{42} R. Petela, Exergy of Heat Radiation, J. Heat Transf., \textbf{86}, 187 (1964). 10.1115/1.3687092

\bibitem{43} R. Petela, Exergy of undiluted thermal radiation, Sol. Energy, \textbf{74}, 469 (2003). 10.1016/S0038-092X(03)00226-3

\bibitem{44} W. Shockley, and H. J. Queisser, Detailed Balance Limit of Efficiency of p-n Junction Solar Cells, J. Appl. Phys., \textbf{32}, 510 (1961). 10.1063/1.1736034

\bibitem{45} Z. Li, and B. Wei, Surpassing Shockley–Queisser Efficiency Limit in Photovoltaic Cells, Nano-micro Lett., \textbf{17}, 330 (2025). 10.1007/s40820-025-01844-8

\bibitem{46} R. M. France, J. F. Geisz, T. Song, W. Olavarria, M. Young, A. Kibbler, and M. A. Steiner, Triple-junction solar cells with 39.5$\%$ terrestrial and 34.2$\%$ space efficiency enabled by thick quantum well superlattices, Joule, 2022 \textbf{6}, 1121-1135. 10.1016/j.joule.2022.04.024

\bibitem{47} J. F. Geisz, R. M. France, K. L. Schulte, M. A. Steiner, A. G. Norman, H. L. Guthrey, M. R. Young, T. Song, and T. Moriarty, Six-junction III–V solar cells with 47.1$\%$ conversion efficiency under 143-Suns concentration, Nat. Energy, 5, 326 (2020). 10.1038/s41560-020-0598-5

\bibitem{48} R. T. Ross, and A. J. Nozik, Efficiency of hot-carrier solar energy converters, J. Appl. Phys., \textbf{53}, 3813 (1982). 10.1063/1.331124

\bibitem{49} A. De Vos, Detailed balance limit of the efficiency of tandem solar cells, J. Phys. D: Appl. Phys., \textbf{13}, 839 (1980). 10.1088/0022-3727/13/5/018

\bibitem{50} R. Loudon, \emph{The Quantum Theory of Light}, (Oxford University Press, 3rd edition, 2000).

\bibitem{51} Y. Candau, On the exergy of radiation, Sol. Energy, \textbf{75}, 241 (2003). 10.1016/j.solener.2003.07.012

\bibitem{52} E. Boukobza, and D. J. Tannor, Entropy exchange and entanglement in the Jaynes-Cummings model, Phys. Rev. A, \textbf{71}, 063821 (2005). 10.1103/PhysRevA.71.063821

\bibitem{53} T. T. Tsutsui, D. Cius, A. Vidiella-Barranco, A. S. M. de Castro, and F. M. Andrade, Revisiting the Jaynes–Cummings Model with Time-dependent Coupling, Braz. J. Phys., \textbf{56}, 21 (2026). 10.1007/s13538-025-01949-w

\bibitem{54} S. Bose, I. Fuentes-Guridi, P. L. Knight, and V. Vedral, Subsystem Purity as an Enforcer of Entanglement, Phys. Rev. Lett., \textbf{87}, 050401 (2001). 10.1103/PhysRevLett.87.050401

\bibitem{55} S. J. D. Phoenix, and P. L. Knight, Fluctuations and entropy in models of quantum optical resonance, Ann. Phys. (N. Y.), \textbf{186}, 381 (1988). 10.1016/0003-4916(88)90006-1

\bibitem{56} Y. N. Zhou, L. Mao, and H. Zhai, Rényi entropy dynamics and Lindblad spectrum for open quantum systems, Phys. Rev. Res., \textbf{3}, 043060 (2021). 10.1103/PhysRevResearch.3.043060

\bibitem{57} K. Kobayashi, Time evolution of the von Neumann entropy in open quantum system, ArXiv:2405.11824 (2024).  10.48550/arXiv.2405.11824

\bibitem{58} J. A. Anaya-Contreras, H. M. Moya-Cessa, and A. Zúñiga-Segundo, The von Neumann Entropy for Mixed States, Entropy, \textbf{21}, 49 (2019). 10.3390/e21010049

\bibitem{59} V. G. Morozov, and G. Röpke, Entropy production in open quantum systems: exactly solvable qubit models, Condens. Matter Phys., \textbf{15}, 43004 (2012). 10.5488/CMP.15.43004

\bibitem{60} M. Brune, F. Schmidt-Kaler, A. Maali, J. Dreyer, E. Hagley, J. M. Raimond, and S. Haroche, Quantum Rabi Oscillation: A Direct Test of Field Quantization in a Cavity, Phys. Rev. Lett., \textbf{76}, 1800 (1996). 10.1103/PhysRevLett.76.1800

\bibitem{61} M. A. D. Taylor, A. Mandal, and P. Huo, Light–matter interaction Hamiltonians in cavity quantum electrodynamics, Chem. Phys. Rev., \textbf{6}, 011305 (2025). 10.1063/5.0225932

\bibitem{62} A. Vaaranta, M. Cattaneo, and P. Muratore-Ginanneschi, Analytical solution of the open dispersive Jaynes-Cummings model and perturbative analytical solution of the open quantum Rabi model, Phys. Rev. A, \textbf{111}, 053717 (2025). 10.1103/PhysRevA.111.053717

\bibitem{63} N. Islam, and S. Biswas, Generalization of the Einstein coefficients and rate equations under the quantum Rabi oscillation, J. Phys. A: Math. Theor., \textbf{54}, 155301 (2021). 10.1088/1751-8121/abeca9

\bibitem{64} J. H. Eberly, N. B. Narozhny, and J. J. Sanchez-Mondragon, Periodic Spontaneous Collapse and Revival in a Simple Quantum Model, Phys. Rev. Lett., \textbf{44}, 1323 (1980). 10.1103/PhysRevLett.44.1323

\bibitem{65} N. Kosugi, S. Matsuo, K. Konno, and N. Hatakenaka, Theory of damped Rabi oscillations, Phys. Rev. B, \textbf{72}, 172509 (2005). 10.1103/PhysRevB.72.172509

\bibitem{66} J. G. Peixoto de Faria, and M. C. Nemes, Dissipative dynamics of the Jaynes-Cummings model in the dispersive approximation: Analytical results, Phys. Rev. A, \textbf{59}, 3918 (1999). 10.1103/PhysRevA.59.3918

\bibitem{67} J. Seke, Extended Jaynes–Cummings model in a damped cavity, J. Opt. Soc. Am. B, \textbf{2}, 1687 (1985). 10.1364/JOSAB.2.001687

\bibitem{68} H. A. Hessian, F. A. Mohammed, and A.-B. A. Mohamed, Entropy and Entanglement in Master Equation of Effective Hamiltonian for Jaynes–Cummings Model, Commun. Theor. Phys., \textbf{51}, 723 (2009). 10.1088/0253-6102/51/4/27

\bibitem{69} K. Hasebe, A. Nakayama, T. Okubo, and M. Yamanoi, Integrodifferential equation in the multimode Jaynes-Cummings model, Phys. Rev. A, \textbf{106}, 062201 (2022). 10.1103/PhysRevA.106.062201

\bibitem{70} J. F. Leandro, and F. L. Semião, Entanglement in Weisskopf–Wigner theory of atomic decay in free space, Opt. Commun., \textbf{282}, 4736 (2009). 10.1016/j.optcom.2009.08.059

\bibitem{71} F. Campaioli, J. H. Cole, and H. Hapuarachchi, Quantum Master Equations: Tips and Tricks for Quantum Optics, Quantum Computing, and Beyond, PRX Quantum, \textbf{5}, 020202 (2024). 10.1103/PRXQuantum.5.020202

\bibitem{72} A. Watanabe, and H. Nakazato, Exact master equation for an open Jaynes–Cummings system, Ann. Phys. (N. Y.), \textbf{441}, 168890 (2022). 10.1016/j.aop.2022.168890

\bibitem{73} J. C. C. Capella, A. Fonseca, P. L. Saldanha, and D. Felinto, High-entanglement regimes in the Weisskopf-Wigner theory for spontaneous decay, Phys. Rev. A, \textbf{112}, 052201 (2025). 10.1103/wmy8-jpvp

\bibitem{74} M. Esposito, K. Lindenberg, and C. Van den Broeck, Entropy production as correlation between system and reservoir, New J. Phys., \textbf{12}, 013013 (2010). 10.1088/1367-2630/12/1/013013

\bibitem{75} A. S. Trushechkin, On the General Definition of the Production of Entropy in Open Markov Quantum Systems, J. Math. Sci., \textbf{241}, 191 (2019). 10.1007/s10958-019-04417-4

\bibitem{76} H. Spohn, Entropy production for quantum dynamical semigroups, J. Math. Phys., \textbf{19}, 1227 (1978). 10.1063/1.523789

\bibitem{77} Md. M. Ali, W.-M. Huang, and W.-M. Zhang, Quantum thermodynamics of single particle systems, Sci. Rep., \textbf{10}, 13500 (2020). 10.1038/s41598-020-70450-y

\bibitem{78} R. H. Lehmberg, Radiation from an N-Atom System. II. Spontaneous Emission from a Pair of Atoms, Phys. Rev. A, \textbf{2}, 889 (1970). 10.1103/PhysRevA.2.889

\bibitem{79} T. M. Clarke, and J. R. Durrant, Charge Photogeneration in Organic Solar Cells, Chem. Rev., \textbf{110}, 6736 (2010). 10.1021/cr900271s

\bibitem{80} S. Datta, and X. Marie, Excitons and excitonic materials, MRS Bull., \textbf{49}, 852 (2024). 10.1557/s43577-024-00766-x

\bibitem{81} A. Chernikov, T. C. Berkelbach, H. M. Hill, A. Rigosi, Y. Li, B. Aslan, D. R. Reichman, M. S. Hybertsen, and T. F. Heinz, Exciton Binding Energy and Nonhydrogenic Rydberg Series in Monolayer
WS$_2$, Phys. Rev. Lett., \textbf{113}, 076802 (2014). 10.1103/PhysRevLett.113.076802

\bibitem{82} T. C. Doan, J. Li, J. Y. Lin, and H. X. Jiang, Bandgap and exciton binding energies of hexagonal boron nitride probed by photocurrent excitation spectroscopy, Appl. Phys. Lett., \textbf{109}, 122101 (2016). 10.1063/1.4963128

\bibitem{83} A. Sugie, K. Nakano, K. Tajima, I. Osaka, and H. Yoshida, Dependence of Exciton Binding Energy on Bandgap of Organic Semiconductors, J. Phys. Chem. Lett., \textbf{14}, 11412 (2023). 10.1021/acs.jpclett.3c02863

\bibitem{84} N. R. Monahan, K. W. Williams, B. Kumar, C. Nuckolls, and X.-Y. Zhu, Direct Observation of Entropy-Driven Electron-Hole Pair Separation at an Organic Semiconductor Interface, Phys. Rev. Lett., \textbf{114}, 247003 (2015). 10.1103/PhysRevLett.114.247003

\bibitem{85} B. Qi, and J. Wang, Open-circuit voltage in organic solar cells, J. Mater. Chem., \textbf{22}, 24315 (2012). 10.1039/C2JM33719C

\bibitem{86} J. Wang, Open-circuit voltage, fill factor, and heterojunction band offset in semiconductor diode solar cells, EcoMat, \textbf{4(6)}:e12263 (2022). 10.1002/eom2.12263

\bibitem{87} N. K. Elumalai, and A. Uddin, Open circuit voltage of organic solar cells: an in-depth review, Energy Environ. Sci., \textbf{9}, 391 (2016). 10.1039/C5EE02871J

\bibitem{88} L. J. A. Koster, V. D. Mihailetchi, R. Ramaker, and P. W. M. Blom, Light intensity dependence of open-circuit voltage of polymer:fullerene solar cells, Appl. Phys. Lett., \textbf{86}, 123509 (2005). 10.1063/1.1889240

\bibitem{89} J. C. Blakesley, and D. Neher, Relationship between energetic disorder and open-circuit voltage in bulk heterojunction organic solar cells, Phys. Rev. B, \textbf{84}, 075210 (2011). 10.1103/PhysRevB.84.075210

\bibitem{90} E. O. Shalenov, Y. S. Seitkozhanov, C. Valagiannopoulos, A. Ng, K. N. Dzhumagulova, and A. N. Jumabekov, Performance evaluation of different designs of back-contact perovskite solar cells, Sol. Energy Mater. Sol. Cells, \textbf{234} 111426 (2022). 10.1016/j.solmat.2021.111426

\bibitem{91} S. M. Ross, \emph{Introduction to Probability Models} (Academic Press, 10th edition, 2010).

\bibitem{92} R. Ravichandran, A. X. Wang, and J. F. Wager, Solid state dielectric screening versus band gap trends and implications, Opt. Mater., \textbf{60}, 181 (2016). 10.1016/j.optmat.2016.07.027

\bibitem{93} K. Yim, Y. Yong, J. Lee, K. Lee, H.-H. Nahm, J. Yoo, C. Lee, C. S. Hwang, and S. Han, Novel high-k dielectrics for next-generation electronic devices screened by automated ab initio calculations, NPG Asia Mater., \textbf{7}, e190 (2015). 10.1038/am.2015.57

\bibitem{94} S. Buckley, K. Rivoire, and J. Vuckovic, Engineered quantum dot single-photon sources, Rep. Prog. Phys., \textbf{75}, 126503 (2012). 10.1088/0034-4885/75/12/126503

\bibitem{95} X. Zianni, and A. G. Nassiopoulou, Photoluminescence lifetimes of Si quantum dots, J. Appl. Phys., \textbf{100}, 074312 (2006). 10.1063/1.2356907

\bibitem{96} S. Asahi, H. Teranishi, K. Kusaki, T. Kaizu, and T. Kita, Two-step photon up-conversion solar cells, Nat. Commun., 8, 14962 (2017). 10.1038/ncomms14962

\bibitem{97} Y. P. Varshni, Temperature dependence of the energy gap in semiconductors, Physica, \textbf{34}, 149 (1967). 10.1016/0031-8914(67)90062-6


\end{thebibliography}
\end{document}